\documentstyle[twocolumn,aps]{revtex}
\begin{document}
\draft
\title{
Functional Bosonization of Interacting Fermions in Arbitrary Dimensions}
\author{Peter Kopietz and Kurt Sch\H{o}nhammer}
\address{
Institut f\H{u}r Theoretische Physik der Universit\H{a}t G\H{o}ttingen,\\
Bunsenstr.9, D-37073 G\H{o}ttingen, Germany}
\date{July 11, 1994}
\maketitle
\begin{abstract}
We bosonize the long-wavelength
excitations
of interacting fermions
in arbitrary dimension
by directly applying a
suitable Hubbard-Stratonowich transformation
to the Grassmannian generating functional
of the fermionic correlation functions.
With this technique we derive a
surprisingly simple expression for the
single-particle Greens-function,
which is valid for arbitrary interaction strength
and can describe
Fermi- as well as Luttinger liquids.
Our approach sheds further light on the relation
between bosonization and the random-phase approximation,
and enables us to study screening
in a non-perturbative way.

\end{abstract}
\narrowtext

Several groups\cite{Luther79,Haldane92,Houghton93,Fradkin94}
have recently
constructed bosonization rules for interacting
fermions in dimensions $d > 1$.
The strategy adopted in these works follows closely
the usual bosonization of one-dimensional systems\cite{Haldane81},
and is based on the observation
that suitably defined local density operators
approximately obey bosonic commutation relations
in the Hilbert space
of states with wave-vectors close to the
Fermi surface.
An alternative method to bosonize
interacting fermions
without calculating commutators
is based on functional
integration\cite{Fogedby76,Chen88}.
In this letter we shall
further develop this approach,
and show that it is in many respects more
powerful than the usual operator bosonization.

Consider a system of interacting fermions
on a $d$-dimensional hypercube with volume ${\cal{V}}$. The hamiltonian is
given by $\hat{H} =
\hat{H}_{0} + \hat{H}_{int}$, with
$\hat{H}_{0} = \sum_{\bf{k} } \epsilon_{\bf{k}} \hat{c}^{\dagger}_{\bf{k} }
\hat{c}_{\bf{k} }$ and
$ \hat{H}_{int} = \frac{1}{2 {\cal{V}}}
\sum_{  \bf{q} \bf{k}  \bf{k}^{\prime}   }
 f_{\bf{q}}^{\bf{k}  \bf{k}^{\prime}  }
 \hat{c}^{\dagger}_{\bf{k+q} }
 \hat{c}^{\dagger}_{\bf{k^{\prime}-q}  }
 \hat{c}_{\bf{k}^{\prime} }
 \hat{c}_{\bf{k}  }
$,
where $\hat{c}_{\bf{k} }$ annihilates an electron with wave-vector ${\bf{k}}$,
and $\epsilon_{\bf{k}}$ is some arbitrary energy dispersion\cite{footnotespin}.
We assume that the degrees of freedom far away from the Fermi surface have been
integrated out, so that ${\bf{k}}$ and ${\bf{k}}^{\prime}$ are restricted
to a shell around the Fermi surface, with radial thickness small compared with
the
Fermi wave-vector $k_{F}$.
Thus, $f_{ \bf{q}}^{\bf{ k k^{\prime}}}$ and $\epsilon_{\bf{k}}$
differ from the bare parameters by finite renormalizations.
Following Refs.\cite{Luther79,Haldane92,Houghton93,Fradkin94},
we partition the Fermi surface
into disjoint patches of volume
$\Lambda^{d-1}$.
The precise shape of the patches is irrelevant as long as we are
interested in fluctuations with wave-vectors $|{\bf{q}}| \ll \Lambda$,
and the interaction
$f_{\bf{q}}^{ \bf{k} \bf{k}^{\prime} }$
is dominated by  momentum transfers
$| {\bf{q}} | \ll k_{F}$.
Throughout this work we shall assume that these restrictions
are satisfied.
Let us
introduce a label $\alpha$ to enumerate the
patches in some arbitrary ordering, and denote by $K^{\alpha}$
the set of ${\bf{k}}$-points
in patch $\alpha$.
For each ${\alpha}$ we define
local density operators
$\hat{R}^{\alpha}_{\bf{q}} = \sum_{\bf{k}} \Theta^{\alpha} ( {\bf{k}} )
\hat{c}^{\dagger}_{\bf{k}} \hat{c}_{\bf{k+q}}$, where
$\Theta^{\alpha} ( {\bf{k} }) = 1 $ if ${\bf{k}} \in K^{\alpha}$,
and zero otherwise.
Assuming that the variations of $f^{ {\bf{k}} {\bf{k}}^{\prime} }_{\bf{q}}$
are negligible if
${\bf{k}}$ and ${\bf{k}}^{\prime}$ are restricted to given patches,
we may introduce the coarse-grained interaction function
 $f^{\alpha \alpha^{\prime}}_{\bf{q}} =
  \ll
  \Theta^{\alpha} ( {\bf{k}} )
  \Theta^{\alpha^{\prime} } ( {\bf{k}}^{\prime} )
  f^{ \bf{k k^{\prime}} }_{\bf{q}}
   \gg
  $
where $\ll \ldots \gg$ denotes averaging with respect to
${\bf{k}}$ and ${\bf{k}}^{\prime}$.

To bosonize ${\hat{H}}$, we consider the imaginary-time correlation function
 \begin{equation}
 \Pi^{\alpha  \alpha^{\prime} }_{ q }
 = \frac{1}{\cal{V}} \int_{0}^{\beta}  d \tau
 e^{ - i \omega_{m}  \tau }
 < \hat{R}^{\alpha}_{\bf{q}} ( \tau )
 \hat{R}^{ \alpha^{\prime} }_{ -{\bf{q} } } ( 0 )  >
 \label{eq:cordef}
 \; \; \; ,
 \end{equation}
where $\beta = 1/T$ is the inverse temperature
and  $ q = [ {\bf{q}} , i \omega_{m} ]$
denotes wave-vector ${\bf{q}}$ and bosonic Matsubara frequency $\omega_{m}
= 2 \pi m T$.
Below it will become evident that
bosonization of $\hat{H}$ is equivalent to the calculation of a functional
$S_{eff} \left\{ \rho \right\}$ of a complex
field $\rho^{\alpha}_{q}$,
such that Eq.\ref{eq:cordef} can be written as
a bosonic functional integral
 \begin{equation}
 \Pi^{\alpha  \alpha^{\prime} }_{ q}
 = \frac{ \beta}{\cal{V}}
 \frac{
\int {\cal{D}} \left\{ \rho \right\}
 \rho^{\alpha}_{ q }
 \rho^{\alpha^{\prime} }_{ -q }
 e^{ -  S_{eff} \left\{ \rho \right\} }
 }
 { \int {\cal{D}} \left\{ \rho \right\} e^{ -  S_{eff} \left\{ \rho
 \right\}  }}
 \label{eq:Seffrho}
 \; \; \; .
 \end{equation}
We now derive $S_{eff} \left\{ \rho \right\}$
in arbitrary dimension.
Starting point is the representation of
$\Pi^{\alpha \alpha^{\prime} }_{q }$
as a functional integral over Grassmann fields
$c_{ k }$\cite{Popov87}
 \begin{equation}
 \Pi^{\alpha  \alpha^{\prime} }_{q }
 = \frac{ \beta}{\cal{V}} \frac{ \int {\cal{D}} \left\{ c \right\}
 R^{\alpha}_{ q }
 R^{\alpha^{\prime} }_{ -q }
 e^{ -  S_{1} \left\{ c \right\} - S_{4} \left\{ c \right\} }
 }
{ \int {\cal{D}} \left\{ c \right\} e^{ -  S_{1} \left\{ c \right\}
 - S_{4} \left\{ c \right\} }}
 \; \; \; ,
 \end{equation}
with $S_{1} \left\{ c \right\} = \beta \sum_{ k }
[ - i \tilde{\omega}_{n} + \xi_{\bf{k}} ] c_{{{k}} }^{\dagger}
c_{{{k}} }$, and
 $
 S_{4} \left\{ c \right\}
  = \frac{\beta}{2 {\cal{V}}} \sum_{q } \sum_{\alpha
\alpha^{\prime} }
 R^{\alpha}_{ -q}
 f_{\bf{q}}^{\alpha  \alpha^{\prime} }
 R^{\alpha^{\prime}}_{ q } $.
Here $\tilde{\omega}_{n} =  \pi  ( 2 n + 1 )T $ are
fermionic Matsubara frequencies,
$\xi_{\bf{k}} = \epsilon_{\bf{k}} - \mu$ is the energy measured
relative to the chemical potential $\mu$ of the
interacting system, and
$R^{\alpha}_{q } = \sum_{ k} \Theta^{\alpha} ( {\bf{k}} )
{c}^{\dagger}_{ {k} } {c}_{k+q}$.
We now decouple $S_{4} \left\{ c \right\}$ by means of a Hubbard-Stratonowich
transformation that involves
two auxiliary fields $\phi^{\alpha}_{q }$ and
$\rho^{\alpha}_{ q}$\cite{Chen88}.
Completing squares and
using the invariance of the integration
measure with respect to shifts\cite{Popov87}, it is not difficult to show that
 \begin{equation}
 e^{  - S_{4} \left\{ c \right\} }
 =
  \int {\cal{D}} \left\{ \rho \right\}
  {\cal{D}} \left\{ \phi \right\}
 e^{  - S_{2} \left\{
 c, \phi  \right\} - S_{3} \left\{ \phi , \rho \right\}
 - S_{int} \left\{ \rho \right\}
 }
 \; \;  ,
 \label{eq:decouple}
 \end{equation}
 \begin{eqnarray}
 S_{2} \left\{ c, \phi  \right\} & = &
 \sum_{k k^{\prime}} c^{\dagger}_{k} V_{k k^{\prime}} c_{k^{\prime}}
 \; \;  ,  \; \;
 V_{k k^{\prime}} = i \sum_{\alpha} \Theta^{\alpha} ( {\bf{k}} )
 \phi^{\alpha}_{k - k^{\prime} }
 \label{eq:S2}
 \\
 S_{3} \left\{ \phi , \rho \right\} & = &
 - i \sum_{q} \sum_{\alpha} \phi^{\alpha}_{-q} \rho^{\alpha}_{q}
 \label{eq:S3}
 \\
 S_{int} \left\{ \rho \right\} & = & \frac{1}{2} \sum_{q} \sum_{\alpha
\alpha^{\prime}}
 \rho^{\alpha}_{-q} \tilde{f}^{\alpha \alpha^{\prime} }_{\bf{q}}
 \rho^{\alpha^{\prime}}_{q}
 \; \;  , \;  \;
\tilde{f}^{\alpha \alpha^{\prime}}_{\bf{q}}
 = \frac{ \beta}{\cal{V}} {f}^{\alpha \alpha^{\prime}}_{\bf{q}}
 \label{eq:Sintrho}
 \; .
 \end{eqnarray}
The field $\rho^{\alpha}_{q}$ is
the bosonized local density field, while
$\phi^{\alpha}_{q}$ is the field dual to $\rho^{\alpha}_{q}$.
Because the composite fermionic field ${R}_{{q}}^{\alpha}$ couples to the
dual field $\phi^{\alpha}_{q}$, the latter must be introduced at least at an
intermediate
step in order to be able to integrate the fermions out
at the very beginning.

Because the effective fermionic action is quadratic, we may now
perform the Grassmann integration, and obtain
 \begin{eqnarray}
 & & \hspace{5mm} \Pi^{\alpha  \alpha^{\prime} }_{q } =
 \nonumber
 \\
 & & \hspace{-0mm}
 \frac{\beta}{\cal{V}}
 \frac{
 \int {\cal{D}} \left\{ \rho \right\}
 {\cal{D}} \left\{ \phi \right\}
 \rho^{\alpha}_{ q }
 \rho^{\alpha^{\prime} }_{ -q }
 e^{ - S_{kin} \left\{ \phi \right\}
 - S_{3} \left\{ \phi , \rho \right\}
 - S_{int} \left\{ \rho  \right\} }
 }
 {  \int {\cal{D}} \left\{ \rho \right\}
  {\cal{D}} \left\{ \phi \right\}
 e^{ - S_{kin} \left\{ \phi \right\} - S_{3} \left\{
 \phi , \rho \right\} - S_{int} \left\{ \rho  \right\} }
 }
  \; ,
 \label{eq:Sphirho}
 \end{eqnarray}
where $S_{kin} \left\{ \phi \right\}$ can be expanded as
 $S_{kin} \left\{ \phi \right\}
 = \sum_{n=1}^{\infty} S_{kin}^{(n)} \left\{ \phi \right\}
 $, with
$ S_{kin}^{(n)} \left\{ \phi \right\} = \frac{1}{n \beta^{n}} Tr \left[
{G}_{0} {V}
 \right]^n$.
The trace is over wave-vector and frequency space,
and
$G_{0}$ and $V$ are infinite matrices, with
$ [ G_{0} ] _{ k  k^{\prime} }
 =  \delta_{ {\bf{k}}, {\bf{k}}^{\prime}  } \delta_{ n , n^{\prime} }
 G_{0} ( k) $, and $G_{0} (k) =
[ i \tilde{\omega}_{n} - \xi_{\bf{k}} ]^{-1}$.
The matrix-elements of ${V}$ are defined in Eq.\ref{eq:S2}.
The subscript on $S_{kin}$ indicates that
this quantity is closely related
to the bosonized kinetic energy. Performing the trace in
$S_{kin}^{(n)} \left\{ \phi \right\}$ yields
 \begin{eqnarray}
 S_{kin}^{(n)} \left\{ \phi \right\} & = &
 \frac{1}{n} \sum_{q_{1}  \ldots q_{n} }
 \sum_{\alpha_{1} \ldots \alpha_{n} }
\delta_{ {\bf{q}}_{1} + \ldots
+ {\bf{q}}_{n} , 0 } \delta_{m_{1} + \ldots + m_{n} , 0}
\nonumber
\\
& \times &
 U^{(n)} (q_1 \alpha_{1}  \ldots q_{n}  \alpha_{n} ) \phi^{\alpha_{1}}_{q_{1}}
\cdots
 \phi^{\alpha_{n}}_{q_{n}}
 \; \; \; ,
 \label{eq:Seffphin}
 \end{eqnarray}
 \begin{eqnarray}
 U^{(n)} (q_1 \alpha_{1}  \ldots q_{n}  \alpha_{n} )
 & = &
 \left( \frac{i}{\beta} \right)^n
 \frac{1}{n!} \sum_{P} \sum_{k}
 G_{0}^{\alpha_{P_1}} (k)
 \nonumber
 \\
 &  & \hspace{-33mm} \times
 G_{0}^{\alpha_{P_2}} ( k + q_{P_1} ) \cdots
 G_{0}^{\alpha_{P_n}} ( k + q_{P_1} + \ldots + q_{P_{n-1}} )
 \; ,
 \label{eq:Uvertex}
 \end{eqnarray}
where $G_{0}^{\alpha} ( k ) = \Theta^{\alpha} ( {\bf{k}} ) G_{0} (k)$.
We have used the invariance of $S_{kin}^{(n)} \left\{ \phi \right\}$
under relabeling of the fields
to symmetrize
the vertices $U^{ (n)}$
with respect to the interchange of any two
labels. The sum $\sum_{P}$ is over the $n!$ permutations of $n$ integers, and
$P_{i}$ denotes the image of $i$ under the permutation.
Note that the vertices $U^{(n)}$ are
uniquely determined by $\xi_{\bf{k}}$.
In
$d > 1$ or for one-dimensional models with non-linear energy dispersion,
$U^{(n)} ( q_{1} \alpha_{1} \ldots q_{n} \alpha_{n} )$
are rather complicated functions
of all external momenta and frequencies.
However, in the
Tomonaga-Luttinger model (TLM)
(i.e. for one-dimensional fermions with
linearized energy dispersion)
all $U^{ (n) }$ with $n \geq 3$ vanish identically.
This is a direct consequence of the
{\it{closed loop theorem}}, which is discussed
and proved in Ref.\cite{Bohr81}.
Note that it is necessary to symmetrize the vertices in order
to apply this theorem.

The first term in the expansion of $S_{kin} \left\{ \phi \right\}$ is
$
 S_{kin}^{(1)} \left\{ \phi \right\}
 = i \sum_{\alpha} N^{\alpha} \phi^{\alpha}_{0}
 $,
where $ N^{\alpha} = \sum_{\bf{k}} \Theta^{\alpha} ( {\bf{k}} ) f (
\xi_{\bf{k}} )$
is the number of occupied states
in patch ${\alpha}$. Here $f( \epsilon ) = [ e^{ \beta \epsilon } +1 ]^{-1}$ is
the Fermi function.
For any finite $q$ the contribution from $S_{kin}^{(1)}$ cancels in
Eq.\ref{eq:Sphirho}.
The second-order term is
 \begin{equation}
 S_{kin}^{(2)} \left\{ \phi \right\}
 = \frac{1}{2} \sum_{q} \sum_{\alpha \alpha^{\prime}}
 \phi^{\alpha}_{-q} U^{(2)} ( -q \alpha, q \alpha^{\prime} )
 \phi^{\alpha^{\prime}}_{q}
 \; \; \; ,
 \label{eq:S2eff}
 \end{equation}
where $U^{(2)}( -q \alpha ,q \alpha^{\prime} ) =
\frac{{\cal{V}}}{\beta}
\Pi_{0}^{\alpha \alpha^{\prime} } (q) $, with
 \begin{eqnarray}
 \Pi_{0}^{\alpha \alpha^{\prime} }
 ( q )
  & =  & \frac{1}{2 \cal{V}} \sum_{\bf{k}}
  [
 \Theta^{\alpha} ( {\bf{k}} ) \Theta^{\alpha^{\prime} }
 ( {\bf{k+q}} )
\frac{ f (  \xi_{\bf{k}} )
 - f (  \xi_{\bf{k+q}} ) }
{ \xi_{\bf{k+q}} - \xi_{\bf{k}} - i \omega_{m} }
 \nonumber
 \\
 &  &  \hspace{10mm}
  +
 \left( \alpha \leftrightarrow \alpha^{\prime} , q \rightarrow -q \right)
 ]
 \label{eq:Pi0}
 \; \; \; .
 \end{eqnarray}
The higher order vertices
describe non-local
interactions between the fields.
In the absence of nesting wave-vectors and
Van Hove singularities, they are non-singular and have a finite limit
if all external momenta and frequencies are set equal to zero\cite{Hertz74}.
In this limit $U^{(n)}$ is
diagonal in all patch-labels, and
is proportional to the
$(n-2)^{nd}$ derivative of the density of states at the
Fermi surface with respect to the chemical potential\cite{Hertz74}.
Note that in the TLM
the density of states is approximated by a constant, so that it is
also from this point of view clear that
$U^{(n)} ( q_{i} = 0 )$ vanishes for $n \geq 3$.
However, the closed loop theorem
is stronger, and
guarantees that the higher-order vertices
vanish even at finite values of the $q_{i}$.

At this point we integrate over the field $\phi$. The effective
action $S_{eff} \left\{ \rho \right\}$ in Eq.\ref{eq:Seffrho} is given by
 \begin{equation}
 S_{eff} \left\{ \rho \right\} = S_{int} \left\{ \rho \right\}
 - \ln \left[ \int {\cal{D}} \left\{ \phi \right\}
  e^{-
 S_{kin} \left\{ \phi  \right\}
 - S_{3} \left\{ \phi , \rho \right\}
 } \right]
 \label{eq:Srhores}
 \; \; \; .
 \end{equation}
Performing the integration in Eq.\ref{eq:Srhores}
perturbatively, we obtain an expansion
$S_{eff} \left\{ \rho \right\} = \sum_{n=1}^{\infty} S_{eff}^{(n)}
\left\{ \rho \right\}$, where  $S_{eff}^{(n)}
\left\{ \rho \right\}$ is of the same form as Eq.\ref{eq:Seffphin}, but with
the $\phi$-fields replaced by the $\rho$-fields, and the vertices
$U^{(n)} ( q_1 \alpha_{1}  \ldots q_{n} \alpha_{n} )$ replaced by
a new set of vertices
$\Gamma^{(n)} ( q_1 \alpha_{1} \ldots q_{n} \alpha_{n} )$.
In general $\Gamma^{(n)}$ contains contributions from
all $U^{(m)} $ with $ m \geq n$.
We now truncate the expansion of $S_{kin} \left\{ \phi \right\}$
at the second order.
In the TLM this approximation is exact,
while in $d>1$ it is expected to be
accurate at high densities and for models where the
higher-order derivatives of the density of states
are small.
The integration in Eq.\ref{eq:Srhores} is then
Gaussian, and we obtain
 \begin{equation}
 S_{eff}^{(2)} \left\{ \rho \right\}
 = \frac{1}{2} \sum_{q} \sum_{\alpha \alpha^{\prime}}
 \rho^{\alpha}_{-q} \left[ \tilde{f}^{\alpha \alpha^{\prime}}_{\bf{q}}
 + {\Gamma}^{(2)} ( -q \alpha , q \alpha^{\prime} ) \right]
\rho^{\alpha^{\prime}}_{q}
 \label{eq:Seffrho2}
 \; \; \; ,
 \end{equation}
where ${\Gamma}^{(2)}$
is the matrix inverse to $U^{(2)}$ in the space spanned
by the patch indices.
Since we are interested in long-wavelength
fluctuations, we may
approximate ${\Gamma}^{(2)}$
by its leading term for small $|{\bf{q}}|$.
At high densities ${U}^{(2)} (-q \alpha , q \alpha^{\prime} ) \propto
\delta^{\alpha
\alpha^{\prime} }$,
because the sum in Eq.\ref{eq:Pi0}
is dominated by momenta ${\bf{k}}$ of the order of $k_{F}$,
so that we may set
$ \Theta^{\alpha} ( {\bf{k}} ) \Theta^{ \alpha^{\prime} }
( {\bf{k+q}} )  \approx
\delta^{\alpha \alpha^{\prime}}
\Theta^{\alpha}
( {\bf{k}} )  $ in Eq.\ref{eq:Pi0}.
In this approximation we obtain
 \begin{equation}
 {\Gamma}^{(2)} ( -q \alpha , q \alpha^{\prime} )
 \approx \delta^{\alpha \alpha^{\prime}} \frac{\beta}{ {\cal{V}} \nu^{\alpha} }
 \frac{  {\xi}^{\alpha}_{\bf{q}} - i \omega_{m} }
 {  {\xi}^{\alpha}_{\bf{q}} }
 \; \; \; ,
 \label{eq:Gammasmall}
 \end{equation}
where $\nu^{\alpha} = {\cal{V}}^{-1} \partial N^{\alpha} / \partial \mu$ is the
local
density of states in patch ${\alpha}$, and
$\xi^{\alpha}_{\bf{q}} = {\bf{v}}^{\alpha} \cdot {\bf{q}}$, with
$ {\bf{v}}^{\alpha} = \nabla_{\bf{k}}
\epsilon_{\bf{k}} |_{ {\bf{k}} \in K^{\alpha} }$.
Inserting Eq.\ref{eq:Gammasmall} into Eq.\ref{eq:Seffrho2}, we see that
the term proportional to $i \omega_{m}$
defines the dynamics of the $\rho$-field.
We now recall that in the
functional integral for canonically quantized bosons
{\it{the coefficient of the term proportional to $-i \omega_{m}$ should be
precisely
$\beta$.}}  Thus, to write our effective action in terms of a canonical
boson field $b_{q}^{\alpha}$, we should rescale the $\rho$-field
accordingly.
This is achieved by substituting
 $\rho^{\alpha}_{q} = ({ {\cal{V}} \nu^{\alpha} |  {\xi}^{\alpha}_{\bf{q}} |
})^{\frac{1}{2}} [
 \Theta ( {\xi}^{\alpha}_{\bf{q}} ) b^{ \alpha}_{q} +
 \Theta ( - {\xi}^{\alpha}_{\bf{q}} ) b^{\dagger \alpha}_{-q}
 ]$ in
Eq.\ref{eq:Seffrho2}.
Our final result for the
bosonized action
$ S^{(2)} \left\{ b \right\} \equiv S_{eff}^{(2)} \left\{ \rho (b) \right\}$
is
 \begin{eqnarray}
 S^{(2)}
 \left\{ b \right\}
 & = &
  \beta \sum_{q} \sum_{\alpha  } \Theta ( \xi^{\alpha }_{\bf{q}} )
  ( - i \omega_{m} )
 b^{ \alpha \dagger }_{q} b^{\alpha}_{q}
 \nonumber
 \\
 & + &
  \beta  \left[ H_{kin}
 \left\{ b \right\}
 +  H_{int}
 \left\{ b \right\} \right]
 \; \; \; ,
 \label{eq:Sb2}
 \end{eqnarray}
 \begin{eqnarray}
 H_{kin} \left\{ b \right\} & = &
   \sum_{q} \sum_{\alpha  } \Theta ( \xi^{\alpha }_{\bf{q}} )
 {\xi}^{\alpha}_{\bf{q}}
 b^{\alpha \dagger }_{q} b^{\alpha}_{q}
 \; \; \; ,
 \label{eq:Hkin}
 \\
 H_{int} \left\{ b \right\} & = &
 \frac{1}{2} \sum_{q} \sum_{\alpha \alpha^{\prime} }
 \Theta ( \xi^{\alpha }_{\bf{q}} )
 \sqrt{ |  \xi^{\alpha}_{\bf{q}} |
 |  \xi^{\alpha^{\prime}}_{\bf{q}} |}
 \nonumber
 \\
 &  & \hspace{-11mm} \times \left[
 \Theta ( \xi^{\alpha^{\prime} }_{\bf{q}} )
 \left(
 \bar{f}_{\bf{q}}^{\alpha \alpha^{\prime} }
 b^{\alpha \dagger }_{q} b^{\alpha^{\prime}}_{q}
 +
 \bar{f}_{\bf{q}}^{\alpha^{\prime} \alpha }
 b^{ \alpha^{\prime} \dagger }_{q} b^{\alpha}_{q}
 \right)
 \right.
 \nonumber
 \\
 &  &  \hspace{-10mm} +
 \left.
 \Theta ( - \xi^{\alpha^{\prime} }_ {\bf{q}} )
 \left(
 \bar{f}_{\bf{q}}^{\alpha \alpha^{\prime} }
 b^{ \alpha \dagger }_{q} b^{ \alpha^{\prime} \dagger }_{-q}
 +
 \bar{f}_{\bf{q}}^{\alpha^{\prime} \alpha }
 b^{\alpha^{\prime}}_{-q} b^{\alpha}_{q}
 \right)
 \right]
 \label{eq:Hint}
 \;  \; ,
 \end{eqnarray}
where $\bar{f}_{\bf{q}}^{\alpha \alpha^{\prime} } =
 \sqrt{ \nu^{\alpha} \nu^{\alpha^{\prime}} } f^{\alpha
\alpha^{\prime}}_{\bf{q}}$ are
dimensionless couplings.
The functional integral for the $b$-field
is now formally identical with a
standard bosonic functional integral.
The corresponding second quantized bosonic hamiltonian
is therefore $\hat{H}^{b} = \hat{H}^{b}_{kin} + \hat{H}^{b}_{int}$,
where $\hat{H}^{b}_{kin}$ and $\hat{H}^{b}_{int}$ are simply
obtained by replacing
the bosonic fields $b^{\alpha}_{q}$
in Eqs.\ref{eq:Hkin},\ref{eq:Hint} by operators
$\hat{b}^{\alpha}_{\bf{q}}$ satisfying
$[ \hat{b}^{\alpha}_{\bf{q}} , \hat{b}^{\alpha^{\prime} \dagger }_{
{\bf{q}}^{\prime} }
 ] = \delta^{ \alpha \alpha^{\prime} } \delta_{ {\bf{q}} , {\bf{q}}^{\prime}
}$.
The resulting $\hat{H}^{b}$ agrees with
the bosonized Hamiltonian derived
in Refs.\cite{Houghton93,Fradkin94}
by means of an operator approach.
{}From our derivation it is obvious that
{\it{bosonization is a non-perturbative but approximate method in }} $d > 1$.
The non-interacting boson-approximation is
only accurate if
the higher order terms $S_{kin}^{(n)} $,
$n \geq 3$, are irrelevant.

$\hat{H}^{b}$  and $S^{(2)}_{eff} \left\{ \rho \right\}$
(Eq.\ref{eq:Seffrho2})
contain the same physical information.
To calculate $\Pi_{q}^{\alpha \alpha^{\prime} }$,
it is
convenient to parametrize the functional integration in terms of the
original density field $\rho^{\alpha}_{q}$.
Inserting Eq.\ref{eq:Seffrho2} into
Eq.\ref{eq:Seffrho} we obtain
 $\Pi^{\alpha \alpha^{\prime} }_{q} = [ [
 {\Pi_{0}}^{-1} ( q ) + {f}_{\bf{q}}
 ]^{-1} ]^{\alpha \alpha^{\prime}}$
where $\Pi_{0}$ and $f$ should be understood as matrices in the
patch labels.
This is nothing but the
random-phase approximation (RPA), which
is known to be exact in the TLM\cite{Dzyaloshinskii74}.
The physical density-density correlation function
is $\sum_{\alpha \alpha^{\prime} }
\Pi^{\alpha \alpha^{\prime} }_{q} $.
In
$d > 1$ the summation over the patches
gives rise to Landau-damping.

Let us now focus on the single-particle
Greens-function $G^{\alpha} ( {\bf{r}} , \tau )$ associated
with patch $\alpha$\cite{footnote1}.
It is well-known\cite{Hertz74} that
the Hubbard-Stratonowich transformation
reduces the calculation of
$G^{\alpha} ( {\bf{r}} , \tau )$
to the problem of calculating the
Greens-function ${\cal{G}}^{\alpha} ( {\bf{r}} ,
{\bf{r}}^{\prime} , \tau , \tau^{\prime}; \left\{ \phi \right\})$
of an effective non-interacting system
in a dynamical random-field $\phi$.
The crucial observation is that
after linearization of the energy dispersion
{\it{ ${\cal{G}}^{\alpha}$
can be calculated exactly for
a given realization of $\phi$,
because the differential equation for
${\cal{G}}^{\alpha}$
is first order and can be easily solved}}\cite{Fogedby76}.
The physical Greens-function
is then obtained by averaging
${\cal{G}}^{\alpha}$
over the distribution of the field ${\phi}$.
Within the non-interacting boson approximation,
this average involves a trivial Gaussian integration.
The final result for the real-space imaginary-time
Greens-function is
 $
 G^{\alpha} ( {\bf{r}} , \tau  )
   =
 G^{\alpha}_{0} ( {\bf{r}} , \tau  )
 e^{
 Q^{\alpha}
 ( {\bf{r}}  , \tau  ) }$,
with
 \begin{eqnarray}
 Q^{\alpha}
 ( {\bf{r}} , \tau ) & = &
 R^{\alpha}
 - S^{\alpha} ( {\bf{r}} , \tau )
  \; \;   , \;  \;
 R^{\alpha} =  \lim_{ {\bf{r}} , \tau \rightarrow 0} S^{\alpha} ({\bf{r}} ,
\tau )
 \;  ,
 \label{eq:Qdef}
 \\
 S^{\alpha}
 ( {\bf{r}}  , \tau   )
  & = &
 \frac{1}{\beta {\cal{V}}} \sum_{ q }
 \frac{ \left[ f^{RPA}_{q} \right]^{\alpha \alpha}
  \cos ( {\bf{q}} \cdot  {\bf{r}}
  - {\omega}_{m}  \tau  )
 }
 {
 ( i \omega_{m} - \xi^{\alpha}_{\bf{q}} )^{2 }}
  \label{eq:Sdef}
 \; \; \; .
 \end{eqnarray}
Here
$f^{RPA}_{q} = [ f_{\bf{q}}^{-1}  + \Pi_{0} ( q )  ]^{-1}$
is again a matrix in the patch indices.
Eqs.\ref{eq:Qdef} and \ref{eq:Sdef}
can be used to determine
by direct calculation of the Greens-function
if the system
is a Fermi- or Luttinger liquid,
or perhaps belongs even to a different category.
In Refs.\cite{Fradkin94,Castellani94} the Greens-function is also found
to be of the form $G^{\alpha} = G_{0}^{\alpha} e^{Q^{\alpha}}$.
Note, however, that Ref.\cite{Fradkin94} and Ref.\cite{Castellani94}
give different expressions for $Q^{\alpha}$.
Our $Q^{\alpha}$ resembles more
the result derived by Castellani {\it{et al.}}
via Ward identities,
although it is not precisely the same,
because their $Q^{\alpha}$
depends explicitly on an ultraviolet cutoff.
In our approach all cutoff-dependence is
contained in the local density of states $\nu^{\alpha}$.
Below we show that for a quite general
class of interactions Eq.\ref{eq:Sdef}
depends only
on the global density of states $\nu = \sum_{\alpha} \nu^{\alpha}$,
and is therefore manifestly cutoff-independent.
We now
discuss Eqs.\ref{eq:Qdef} and \ref{eq:Sdef} for
$ {\cal{V}} , \beta \rightarrow \infty$
in some physically interesting cases.
A detailed derivation will be given in a longer
publication.

(a) {\it{Luttinger liquid in $d=1$}}. In one dimension
Eqs.\ref{eq:Qdef} and \ref{eq:Sdef} correctly
reproduce the known results
for the Greens-function of the TLM.
Note that $R^{\alpha}$  and
$S^{\alpha} ( {\bf{r}} , \tau )$ are both
logarithmically divergent
if $f^{\alpha \alpha^{\prime} }_{\bf{q}}$
goes to a finite limit for ${\bf{q}} \rightarrow 0$,
but
$Q^{\alpha} ( {\bf{r}} , \tau )$ is finite.
It seems that
the possibility of expressing
the interaction dependence of $Q^{\alpha} ( {\bf{r}} , \tau )$
exclusively in terms of the RPA-interaction
has not been noticed
in the literature on the TLM.

(b) {\it{Quasi-particle residue}}.
If the integral defining $R^{\alpha}$ exists, the system is
by the usual definition a Fermi liquid.
The quasi-particle residue $Z^{\alpha}$
for wave-vectors
${\bf{k}} \in K^{\alpha}$ is then
$Z^{\alpha} = e^{ R^{\alpha} }$.
{}From Eq.\ref{eq:Sdef} it is not difficult to show that
$R^{\alpha}$ is real and negative, so that $0 < Z^{\alpha} \leq 1$.

(c) {\it{Singular interactions}}.
For interactions of the form
$f^{\alpha \alpha^{\prime} }_{\bf{q}} =  g^2  | {\bf{q}} |^{ - \eta }
e^{-  | {\bf{q}} | / q_{c} }$, $\eta \geq 0$, we find
by simple power counting that
$R^{\alpha}$ exists only for
$\eta < 2 ( d-1)$, in agreement with Ref.\cite{Bares93}.
If in addition $ \frac{d-1}{2} < \eta $,
and if the {\it{screening wave-vector}}
$\kappa = ( \nu g^2 )^{1/ \eta}$
is small compared with $k_{F}$, then
the ultraviolet cutoff $q_{c}$ is not necessary and may be
set to infinity,
because $\kappa$ acts as a natural short-distance cutoff
for all wave-vector integrals.
It is important to stress that
we are not a priori assuming that
the bare interaction is screened.
A straight-forward calculation gives
in this case for the
quasi-particle residue
 \begin{equation}
 Z^{\alpha} = \exp \left[ -
 \frac{r(d, \eta )}{ 2 ( d-1) - \eta }  ( \kappa / k_{F} )^{d-1}
 \right]
 \label{eq:Zres}
 \; \; \; ,
 \end{equation}
where the positive numerical constant
$r ( d , \eta )$
can be explicitly written down
as a one-dimensional integral,
and remains finite at $ \eta = 2 (d-1)$.

(d) {\it{Coulomb interaction in $d > 1$}}.
This is a special case of (c):
$g^2 = s_{d} e^2$,
$\eta = d-1$, and $q_{c} = \infty$.
Here $s_{d}$ is a numerical constant ($s_{2} = 2 \pi$,
$s_{3} = 4 \pi$).
$\kappa = ( \nu s_{d} e^2 )^{ \frac{1}{d-1}}$
can now be identified with the usual
Thomas-Fermi screening wave-vector.
Note that
$\kappa \ll k_{F}$ in the
high-density limit we are considering here,
and that $\eta = d-1$ lies for all $d$
in the regime $\frac{ d-1}{2} < \eta < 2 ( d-1)$
where $\kappa$
acts as an intrinsic ultraviolet cutoff.
We conclude that at high densities the Coulomb gas
is a Fermi liquid in all physical dimensions
$d > 1$.
The quasi-particle residue can be obtained
by setting $\eta = d-1$ in Eq.\ref{eq:Zres}.
If we analytically continue to non-integer
dimensions\cite{Castellani94},
we see that $Z^{\alpha} \ll 1 $
for $d-1 \ll ( \kappa / k_{F} )^{d-1}$.
It can be shown that
$r ( d , d-1) \sim \frac{1}{2} + O (d-1)$ for
small $d-1$, so that
$Z^{\alpha} \propto ( k_{F} / \kappa  )^{1/2} \exp[ - \frac{1}{ 2(d-1) }]$
for $d \rightarrow 1$.

(e) {\it{Quasi-particle damping}}.
If $| R^{\alpha} | < \infty$ and
the integrand in Eq.\ref{eq:Sdef}
is a sufficiently smooth function of $q$,
the Fourier integral theorem implies
$|S^{\alpha} ( {\bf{r}} , \tau ) | < \infty$ for all ${\bf{r}}$ and $\tau$, and
$ \lim_{{\bf{r}} , \tau \rightarrow \infty}
S^{\alpha} ( {\bf{r}} , \tau ) =0$.
The precise way in which
$S^{\alpha} ( {\bf{r}} , \tau )$ vanishes
determines the  quasi-particle damping.
For a conventional Fermi liquid with damping
$\gamma_{\bf{q}}^{\alpha} \propto ( \xi^{\alpha}_{\bf{q}})^2$
the leading term vanishes for large
$ u \equiv
{max} \left\{ | {\bf{r}} | , | {\bf{v}}^{\alpha} \tau | \right\}$
as $u^{-1}$.
For interactions
of the form discussed in (c)
we find that
$S^{\alpha} ( {\bf{r}} , \tau ) \sim c_{1} u^{-(d -1 - \eta / 2)}
+ c_{2} u^{-1}$ as $u \rightarrow \infty$,
where $c_{1}$ and $c_{2}$ are constants.
For $ d   < 2 + \frac{\eta}{2}$ the first term decays slower than the second
one,
and gives rise to anomalously large corrections.
In particular, the Coulomb interaction
satisfies this condition for $d < 3$,
{\it{which includes the physically accessible
case}} $d=2$.
The vanishing of
$S^{\alpha} ( {\bf{r}} , \tau )$ for large
$u$ implies that the effective mass is not renormalized.
Thus, the definition of $\epsilon_{\bf{k}}$
in our original hamiltonian $\hat{H}_{0}$ should take
the effective mass renormalization into account.

(f) {\it{Relevance of transverse hopping}}.
Experiments
probing the Luttinger liquid regime
are performed on anisotropic quasi one-dimensional conductors,
consisting of weakly coupled chains embedded in a
three-dimensional lattice.
If the nature of the {\it{three-dimensional}} interaction
$f^{\alpha \alpha^{\prime} }_{\bf{q}}$
is such that
for vanishing interchain hopping $t_{\bot}$ the
system is a Luttinger liquid,
then for any finite $t_{\bot} $ the quasi-particle residue is finite, so that
strictly speaking these systems
are not Luttinger liquids.
If $t_{\bot}$ is small compared with
the characteristic intra-chain hopping energy $t_{\|}$,
then the quasi-particle residue is proportional to
$( {t_{\bot} } / { t_{\|} } )^{\gamma}$,
where the non-universal exponent $\gamma > 0$
{\it{can be identified with the anomalous dimension of the
Luttinger liquid that would exist for $t_{\bot} = 0$}}\cite{Kopietz94b}.

Finally, we would like to point out that
our approach opens the way for a
systematic calculation of corrections
to the non-interacting boson approximation, which
exist even in $d=1$ if the energy dispersion is not linearized\cite{Haldane81}.
Treating the terms $S_{kin}^{(n)}$ for $n \geq 3$ perturbatively within
one-loop approximation, we find that
for interactions of the form $f_{\bf{q}}^{\alpha \alpha^{\prime}} = f_{0} e^{-
| {\bf{q}} | / q_{c} }$,
$q_{c} \ll k_{F}$,
the non-Gaussian corrections
to the  bosonized hamiltonian are negligible
if the dimensionless parameter
$ A_{0} = q_{c}^d \frac{f_{0}}{ 1 + \nu f_{0}} | \frac{\partial \nu }{ \partial
\mu}|$ is small
compared with unity.
Note that $A_{0}$ vanishes
if the energy dispersion is linearized, because then
the density of states is approximated by a constant independent of $\mu$.

We would like to thank G. E. Castilla and V. Meden for
discussions and comments on the
manuscript.

%
%
%
%              R E F E R E N C E S
%


\begin{thebibliography}{99}
\bibitem{footnote1}
$G^{\alpha} ( {\bf{r}} , \tau ) =
\frac{1}{ \beta {\cal{V}} }
\sum_{k} e^{ i ( {\bf{k}} \cdot {\bf{r}} - \tilde{\omega}_{n} \tau )}
\Theta^{\alpha} ( {\bf{k}} ) G ( {\bf{k}} , i \tilde{\omega}_{n} )$,
where $G ( {\bf{k}} , i \tilde{\omega}_{n} )$ is the
usual Matsubara Greens-function.
Note that the non-interacting function $G^{\alpha}_{0} ( {\bf{r}} ,\tau )$
is proportional to $\delta^{(d-1)} ( {\bf{r}}^{\alpha}_{\bot} )$, where
${\bf{r}}_{\bot}^{\alpha}$ is the $d-1$ dimensional projection
of ${\bf{r}}$ perpendicular to ${\bf{v}}^{\alpha}$.
Therefore we may set ${\bf{r}} = x {\bf{v}}^{\alpha}/ | {\bf{v}}^{\alpha}|$ in
Eq.\ref{eq:Sdef}.
\end{thebibliography}
\end{document}